\documentclass[aps,prl,reprint,superscriptaddress,citesort,nolongbibliography,floatfix]{revtex4-2}

\usepackage{graphicx}
\usepackage{amsmath}
\usepackage{overpic}
\usepackage{amsfonts}
\usepackage{tikz}
\usepackage{enumitem}
\usetikzlibrary{backgrounds,arrows.meta,decorations.pathreplacing,positioning,calc,3d}

\usepackage{tabularx}
\usepackage{booktabs,threeparttable}

\usepackage{bm}
\usepackage{lipsum}
\usepackage{hyperref}
\usepackage{xcolor}
\definecolor{tab_blue}{HTML}{1F77B4}
\hypersetup{ breaklinks=true, colorlinks=true, linkcolor=tab_blue, filecolor=tab_blue, urlcolor=tab_blue,
 citecolor=tab_blue,
}

\begin{document}

\title{ Flow-driven hysteresis in the transition boiling regime}

\author{Alessandro Gabbana}
\affiliation{Computational Physics and Methods Group (CCS-2), Los Alamos National Laboratory, Los Alamos, 87545 New Mexico,
 USA}
\affiliation{Center for Nonlinear Studies (CNLS), Los Alamos National Laboratory, Los Alamos, 87545 New Mexico, USA}
\affiliation{University of Ferrara and INFN Ferrara, 44122 Ferrara, Italy}

\author{Xander M. de Wit}
\affiliation{Fluids and Flows group and J.M. Burgers Center for Fluid Mechanics, Eindhoven University of Technology,
 5600 MB Eindhoven, Netherlands}
\affiliation{Computational Physics and Methods Group (CCS-2), Los Alamos National Laboratory, Los Alamos, 87545 New Mexico,
 USA}

\author{Linlin Fei}
\email{linlinfei@xjtu.edu.cn}
\affiliation{Key Laboratory of Thermo-Fluid Science and Engineering of Ministry of Education, School of Energy and 
 Power Engineering, Xi’an Jiaotong University, Xi’an, Shaanxi 710049, China}

\author{Ziqi Wang}
\affiliation{Fluids and Flows group and J.M. Burgers Center for Fluid Mechanics, Eindhoven University of Technology,
 5600 MB Eindhoven, Netherlands}

\author{Daniel Livescu}
\affiliation{Computational Physics and Methods Group (CCS-2), Los Alamos National Laboratory, Los Alamos, 87545 New Mexico,
 USA}

\author{Federico Toschi}
\affiliation{Fluids and Flows group and J.M. Burgers Center for Fluid Mechanics, Eindhoven University of Technology,
 5600 MB Eindhoven, Netherlands}
\affiliation{CNR-IAC, I-00185 Rome, Italy}

\begin{abstract} 
  Transition boiling is an intermediate regime occurring between nucleate boiling, where bubbles at the surface 
  efficiently carry heat away, and film boiling, where a layer of vapor formed over the surface
  insulates the system reducing heat transfer. This regime is inherently unstable and typically occurs near 
  the boiling crisis, where the system approaches the maximum heat flux.
  Transition boiling hysteresis remains a central open problem in phase-change heat
  transfer, with critical implications for industrial cooling systems and nuclear reactor safety, since 
  entering this regime sharply reduces heat removal potentially leading to overheating or component damage.
  
  We investigate the mechanisms driving hysteresis in the transition boiling regime through large-scale 
  three-dimensional numerical simulations, providing clearcut evidence that hysteresis occurs even under 
  idealized conditions of pool boiling on flat surfaces at constant temperature. 
  This demonstrates that hysteresis arises purely from the flow dynamics of the liquid-vapor system, 
  rather than from surface properties or defects.
  Moreover, we disclose strong asymmetries in the transition dynamics between nucleate and film boiling. 
  During heating, the transition is abrupt and memoryless, whereas, upon decreasing 
  the surface temperature, it is more complex, with the emergence of metastable 
  coexisting states that can delay the transition.

\end{abstract}

\maketitle  

\begin{figure*}
    \centering
    \begin{overpic}[width=\linewidth]{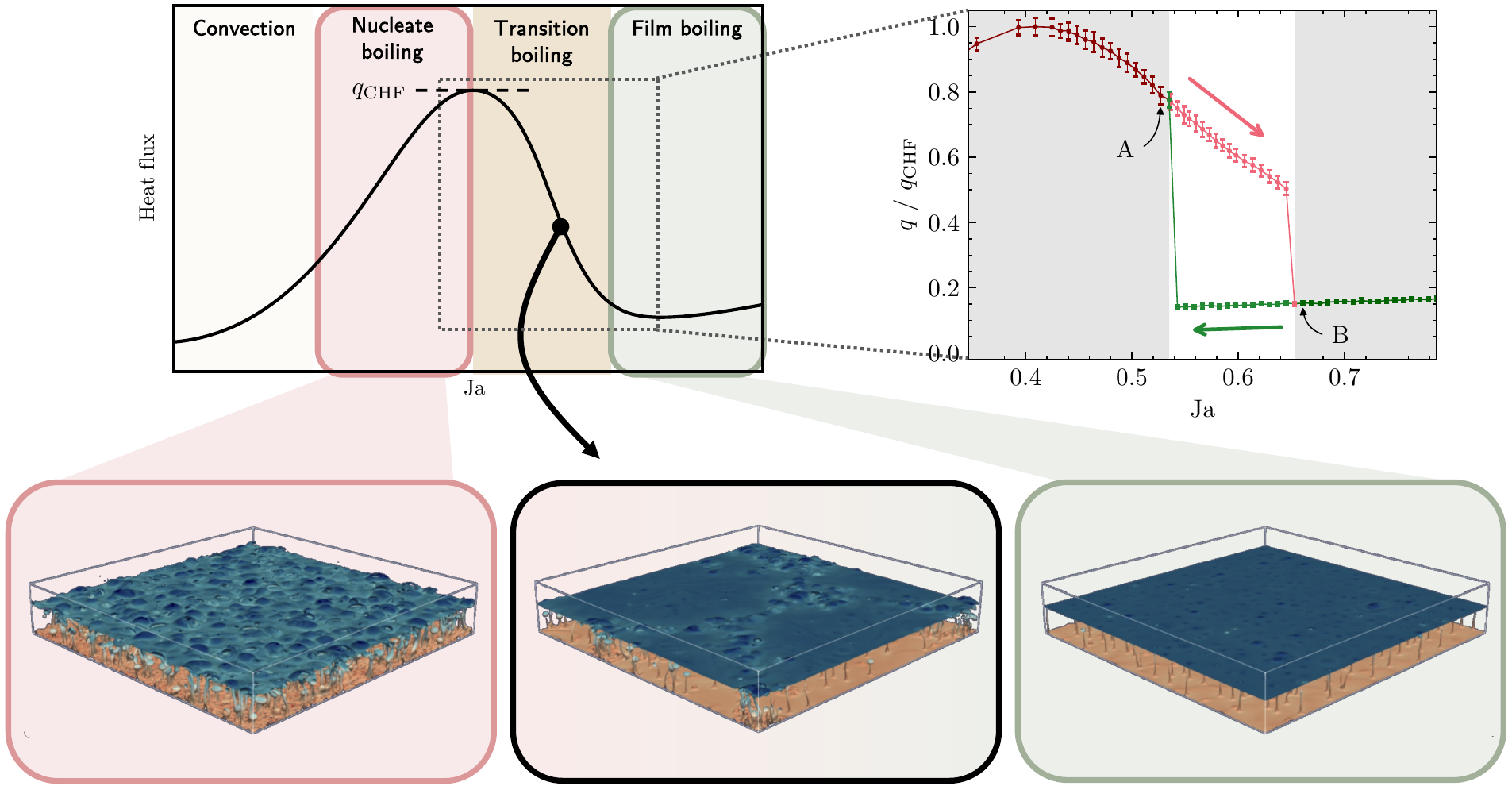}
     \put( 5.5,  51.2){\textbf{ (a) }}
     \put(56  ,  51.2){\textbf{ (b) }}
     \put(1.2 ,  17.5){\textbf{ (c) }}
     \put(34.6,  17.5){\textbf{ (d) }}
     \put(68  ,  17.5){\textbf{ (e) }}
    \end{overpic}
    \caption{ 
      Boiling regimes in pool boiling. (a) Boiling curve, showing the heat flux as a function of Jakob number ($\textrm{Ja}$).
      (b) Detailed view of the transition boiling regime. (c-e) Representative flow states: 
      nucleate boiling (c), metastable coexistence of nucleate and film boiling (d), and film boiling (e). 
      The transition from nucleate to film boiling exhibits pronounced hysteresis, with metastable coexistence 
      observed along the rewetting branch, where the bottom wall temperature is gradually reduced.
    }\label{fig:overview}
\end{figure*}
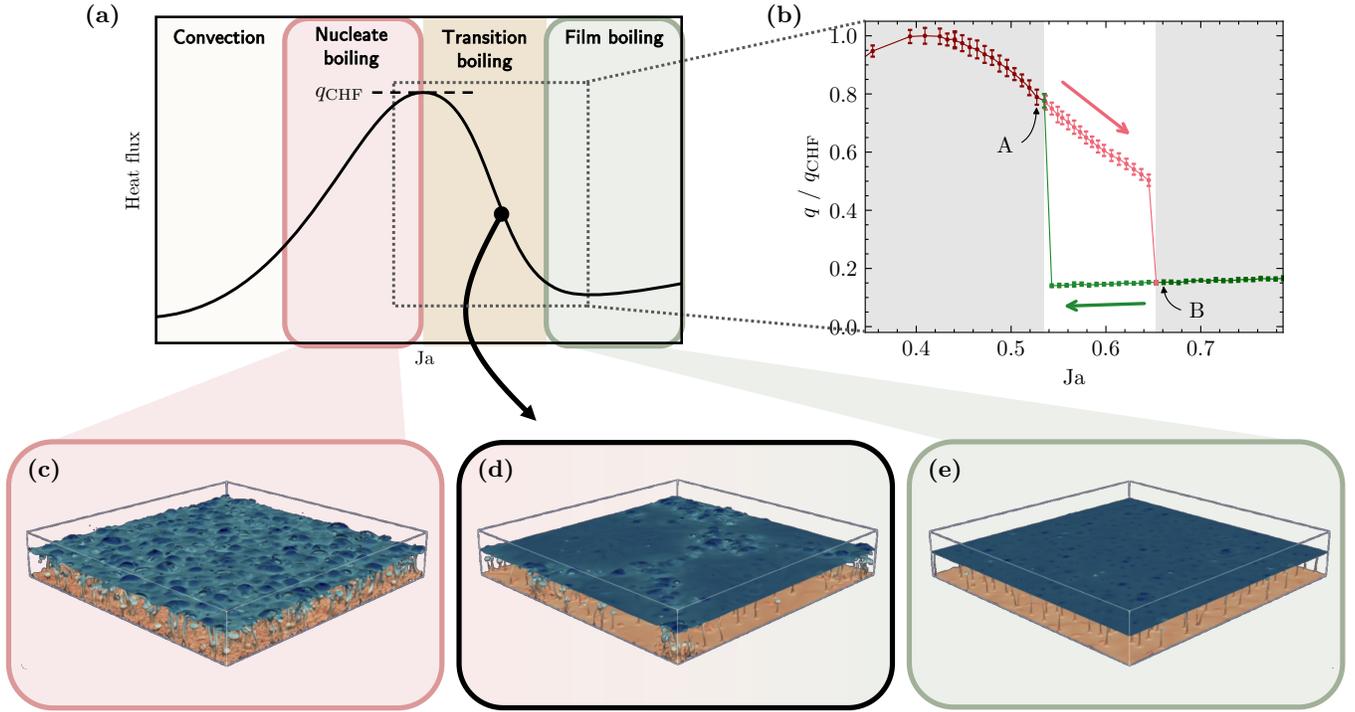

{\it Introduction.---}
Boiling heat transfer is a fundamental and widely used thermal phenomenon in modern engineering systems,
as it offers a very efficient mechanism for extracting and transporting heat.
Its applications span across diverse industries, from nuclear power generation and chemical processing to
electronics cooling, and even aerospace propulsion systems~\cite{ding2025review,benam2021review,ding2016application,
mashaei2015effect,giustini2020modelling,mi2024research,karayiannis2017flow}. 
The phenomenological description of the boiling process is commonly summarized by the boiling curve, i.e., 
the heat flux removed from a heated surface versus the wall superheat, defined as the difference between the surface 
temperature and the saturation temperature of the liquid. %
As shown in Fig.~\ref{fig:overview}a, three distinct regimes can be identified: (i) nucleate
boiling, where discrete bubbles form and detach from the heated surface; (ii) transition boiling, an intermediate regime,
characterized by a decreasing efficiency of the heat transfer; (iii) and film boiling, where a stable vapor film covers
the entire surface~\cite{incropera1985introduction,collier1994convective}. 

Many industrial systems operate near the peak of the boiling curve (the critical heat flux $q_{\rm CHF}$), 
where nucleate boiling provides maximum heat removal. 
Operating too close to this peak can lead to the unstable transition boiling regime,
where heat transfer degrades rapidly and control becomes difficult. This regime is inherently unstable due to the
inverse relationship between surface temperature and heat flux~\cite{dhir1998boiling}, and can occur transiently during
rewetting from film boiling after a boiling crisis. Understanding transition boiling is therefore essential for safe
cooling system operation~\cite{benam2021review}, particularly in nuclear reactors, where departure from nucleate
boiling can damage fuel cladding and create safety hazards~\cite{giustini2020modelling}.

The presence of hysteresis in the transition boiling regime is among the most important and not yet fully understood
problems in boiling heat transfer research~\cite{dhir1998boiling, Ghiaasiaan2017}. Hysteresis refers to the
path-dependent nature of the transition, where the heat flux in the transition boiling regime depends on whether it is
approached from low-temperature nucleate boiling side, when gradually increasing the temperature, or from the
high-temperature film boiling side, when rewetting. 

Experimental investigations into hysteresis in transition boiling have been inconclusive. While hysteresis
was observed in the experiments by Witte \& Lienhard~\cite{witte1982existence} and later
confirmed by other experiments~\cite{Bui1985,Ramilison1987,liaw1986effect}, no such behavior was observed in
several more recent experimental studies~\cite{hohl2001model,auracher2002experimental}. These conflicting observations have been surmised to be related to transient effects 
or impurities on the heated surface~\cite{blum1996stability,hohl2001model,auracher2002experimental,ohtake2006derivations}. 
Moreover, experimental approaches face inherent limitations in providing detailed spatially and temporally resolved measurements of the
microscale phenomena occurring during these transitions~\cite{zhang2019percolative}. These challenges underscore the
need for numerical simulations under idealized conditions to further investigate the presence and mechanisms of
hysteresis in transition boiling~\cite{jiang2023review}.

In recent years, novel mesoscopic approaches have opened up new avenues for modeling liquid-vapor 
phase transition phenomena~\cite{zhang2003lattice,biferale2012convection,gong2012lattice,li2015lattice,huang2021mesoscopic}.
Nevertheless, current numerical capabilities remain largely restricted to two-dimensional domains or
three-dimensional simulations with limited numbers of bubbles and relatively small computational domains.
In this letter, we present a comprehensive numerical investigation of transition boiling hysteresis,
providing evidence of hysteresis in the transition boiling regime under 
idealized conditions of pool boiling on flat surfaces with constant temperature. Furthermore, we observe 
a distinct asymmetry in the transition behavior depending on whether the bottom wall temperature is increased or reduced, 
with more abrupt transitions during heating and more complex, delayed transitions upon reduction of the bottom wall temperature, 
where metastable coexisting states are found to emerge.

{\it Computational Model.---}
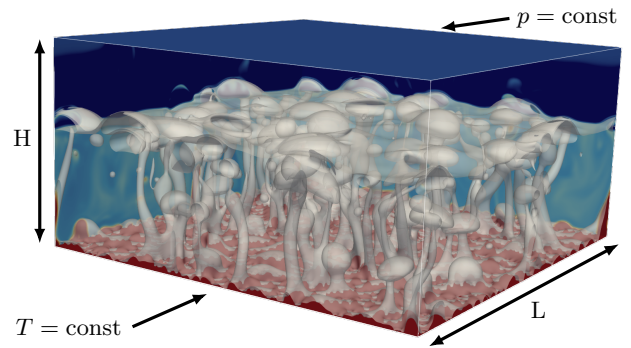
\begin{figure}[b]
    \centering
   \begin{tikzpicture}
    \node[anchor=south west, inner sep=0] (img) {\includegraphics[width=.89\columnwidth]{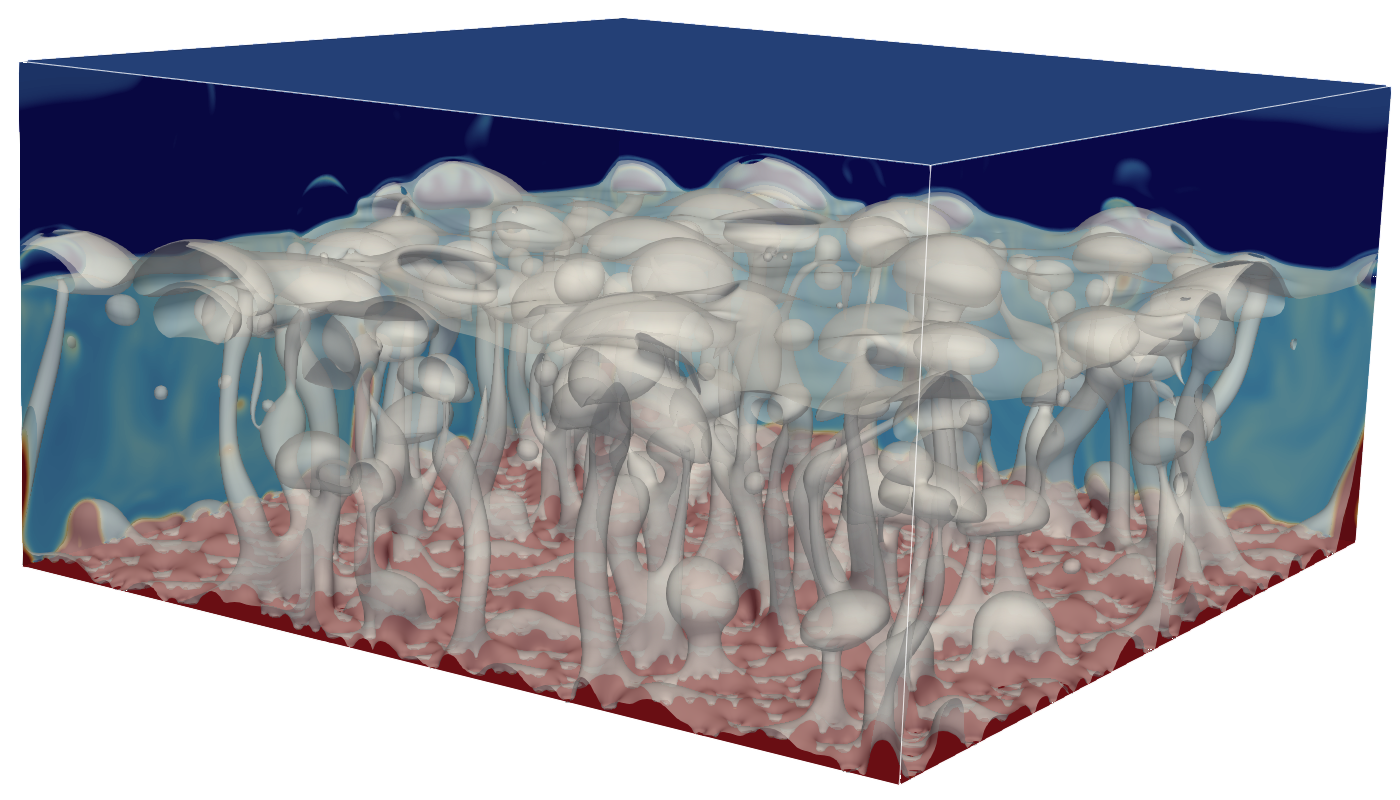}};
    
    \begin{scope}[x={(img.south east)}, y={(img.north west)}]

      \draw[black, very thick, {Latex[length=2mm]}-{Latex[length=2mm]}] (-0.01,0.30) -- (-0.01,0.92);

      \draw[black, very thick, {Latex[length=2mm]}-{Latex[length=2mm]}] (0.66, -0.01) -- (0.99,0.30);

      \draw[black, very thick, -{Latex[length=2mm]}] (0.15, 0.05) -- (0.28,0.15);

      \draw[black, very thick,  {Latex[length=2mm]}-] (0.68, 0.95) -- (0.8,0.98);

      \node[inner sep=2pt, font=\small] at (-0.04,0.62) {H};
      \node[inner sep=2pt, font=\small] at ( 0.85,0.1) {L};

      \node[inner sep=2pt, font=\small] at ( 0.04,0.05) {$T = \textrm{const}$};
      \node[inner sep=2pt, font=\small] at ( 0.90,0.98) {$p = \textrm{const}$};

    \end{scope}
  \end{tikzpicture}
  \caption{Schematic of the computational domain of size $L \times L \times H$, with 
           superheated bottom wall kept at constant temperature and no-slip boundary conditions for the velocity.
           The outflow top boundary is kept at constant pressure, and periodic vertical sides.
          }\label{fig:domain_sketch}
\end{figure}
The time evolution of a boiling system involving a viscous two-phase flow,
with liquid-vapour phase change, is governed by the mass, momentum and energy conservation equations.
The fluid dynamics are described by the compressible Navier–Stokes equations:
\begin{subequations}\label{eq:ns}
  \begin{align}
      \partial_t \rho + \bm{\nabla} \cdot (\rho \bm{u}) &= 0, \\
      \partial_t (\rho \bm{u}) + \bm{\nabla} \cdot (\rho \bm{u} \bm{u}) 
      &= 
      - \bm{\nabla} (\rho c_s^2) 
      + \bm{\nabla} \cdot [ \rho \nu (\bm{\nabla} \bm{u} 
      + (\bm{\nabla} \bm{u})^{\top}) \nonumber\\
      &- \frac{2}{3} \rho \nu (\bm{\nabla} \cdot \bm{u}) \mathbf{I} ] 
      + \bm{\nabla} \left[ \rho \xi (\bm{\nabla} \cdot \bm{u}) \right] 
      + \bm{\Xi},
  \end{align}
\end{subequations} 
where $\rho$ is the local density of the fluid, $c_s$ the speed of sound, $\nu$ and $\xi$ the kinematic and bulk
viscosity, respectively, and $\bm{\Xi}$ denotes the forces due to a pseudo-potential model (see End Matter). 
The temperature field $T$ evolves according to the energy conservation equation~\cite{anderson1998diffuse,hazi2008modeling,li2015lattice}:
\begin{equation}\label{eq:energy}
  \frac{{\partial T}}{{\partial t}} 
  = 
  - {\bm{u}} \cdot \bm{\nabla} T 
  + \bm{\nabla} \cdot (\lambda \bm{\nabla} T) 
  - \frac{T}{{\rho c_v}} \left( \frac{{\partial p_{\rm EOS}}}{{\partial T}} \right)_{\!\rho} \bm{\nabla} \cdot {\bm{u}},
\end{equation} 
where $\lambda$ denotes the conductivity and $c_v$ is the heat capacity. 
In this work, the pressure $p_{\rm EOS}$ is computed from the Peng-Robinson equation of state:
\begin{equation}\label{eq:peng-eos} 
  p_{\rm EOS} = \frac{\rho RT}{1 - b\rho} - \frac{a \varphi(T) \rho^2}{1 + 2b\rho - b^2 \rho^2},
\end{equation} 
where the temperature-dependent function $ \varphi(T) $ and its parameters are defined 
in Refs.~\cite{yuan2006equations,li2013lattice}. See the End Matter for the parameter values used in this work.

We solve Eq.~\eqref{eq:ns} employing the Lattice-Boltzmann method (LBM)~\cite{succi-book-2018}.
The phase change dynamics are captured using a pseudo-potential model~\cite{shan1993lattice,shan1994simulation},
which naturally handles the liquid-vapor interface without requiring explicit interface tracking. 
The scalar energy equation (Eq.~\eqref{eq:energy}) is discretized using a second order finite difference scheme
and time integrated with a standard fourth-order Runge-Kutta scheme.
The numerical framework is described and validated in detail in Ref.~\cite{linlin-pof-2020}, and 
has been recently further developed to support massively parallel execution on multi-GPU clusters, 
thereby enabling large-scale 3D simulations~\cite{eurohpc-proceedings}.

We consider the canonical pool boiling configuration, which is resolved by a  computational domain of size $L \times L \times H$ (Fig.~\ref{fig:domain_sketch}). 
The bottom wall simulates an overheated surface, maintained at a constant temperature 
$T_{\rm{wall}} = T_{\rm{sat}} + \Delta T$ to drive bubble nucleation, with no-slip velocity boundary conditions.
An outflow condition is applied at the top boundary for the velocity field, coupled with constant pressure conditions 
to allow vapour outflow without spurious reflections into the computational domain. 
Periodic boundary conditions are enforced on the remaining four vertical faces.

We establish a mapping between real physical parameters and their corresponding numerical units
by ensuring consistency for the most relevant dimensionless quantities (listed in Table.~\ref{tab:params}). 

\renewcommand{\arraystretch}{1.5} 
\begin{table}[h]
\begin{threeparttable}
\caption{ Summary of the dimensionless parameters used in the simulations. 
          The characteristic velocity is defined as the buoyancy velocity, 
          $v_{B}=\sqrt{g \beta \Delta T H}$.
        }\label{tab:params} 
\begin{ruledtabular}
\begin{tabular*}{0.95\linewidth}{@{\extracolsep{\fill}}lcc}
  Dimensionless quantity & Definition & Value \\
  \midrule
  Prandtl number ($\mathrm{Pr}$)  & $\nu/\kappa$                          & 21.4 \\
  Jakob number ($\mathrm{Ja}$)    & $C_{p}\,\Delta T / h_{fg}$            & 0.4 -- 0.8 \\
  Rayleigh number ($\mathrm{Ra}$) & $g \beta \Delta T H^3 / (\nu \kappa)$ & $(9.15 \text{ -- } 18.31) \times 10^{6}$ \\
  Weber number ($\mathrm{We}$)    & $\rho v_{B}^2 L / \sigma$             & 30.2 -- 60.5 \\
  Froude number ($\mathrm{Fr}$)   & $v_{B} / \sqrt{gH}$                   & 1.02 -- 1.46 \\
\end{tabular*}
\end{ruledtabular}
\end{threeparttable}
\end{table}

The key dimensionless parameter governing the boiling process is the Jakob number:
\begin{equation} 
  \mathrm{Ja} = \frac{c_p \Delta T}{h_{fg}},
\end{equation} 
where %
$\Delta T$ is the wall superheat, and $h_{fg}$ is the latent heat. 
We also define the dimensionless heat flux as $q^{*} = q/q_{\rm CHF}$, where $q_{\rm CHF}$ is the critical heat flux.

{\it Numerical Results.---}
We consider simulations of pool boiling on domains with lateral size ranging from $L = 348$ to $L = 4096$, 
keeping the vertical extent fixed at $H = 512$.  
The grid resolution ensures that for the parameters listed in Table~\ref{tab:params} the bubble departure diameter 
is resolved by approximately 20 grid points, providing adequate resolution for interfacial dynamics.
\begin{figure}[b]
    \centering
    \begin{overpic}[width=0.99\columnwidth]{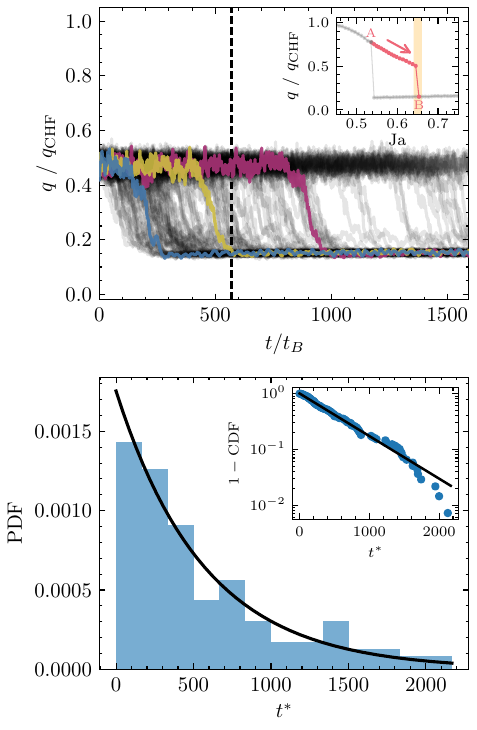}
     \put(-1,  98.0){\textbf{ (a) }}
     \put(-1,  49.0){\textbf{ (b) }}
    \end{overpic}    
    \caption{
    (a) Time evolution of the average heat flux for 80 independent
     realizations restarted from uncorrelated snapshots at $\rm{Ja}=0.527$ (point A in the inset) 
     and run at $\rm{Ja}=0.653$ (point B in the inset). We highlight with colors three different realizations as an example.
     The crossover time $t^{*}$ is defined as the time required for the heat flux to drop to within $10\%$ of 
     its mean film boiling value. The average value of this quantity, $\mu \approx 570$ in the units 
     of the characteristic buoyancy time $t_{B} = H/v_{B}$, is shown by the vertical dotted line. %
     (b) Probability density function of $t^{*}$, the crossover time from transitional to film boiling, together with a
     fit to an exponential distribution (black line). 
     }\label{fig:small_system_analysis} 
\end{figure}
\begin{figure*}[htb]
    \centering
    \includegraphics[width=0.99\linewidth]{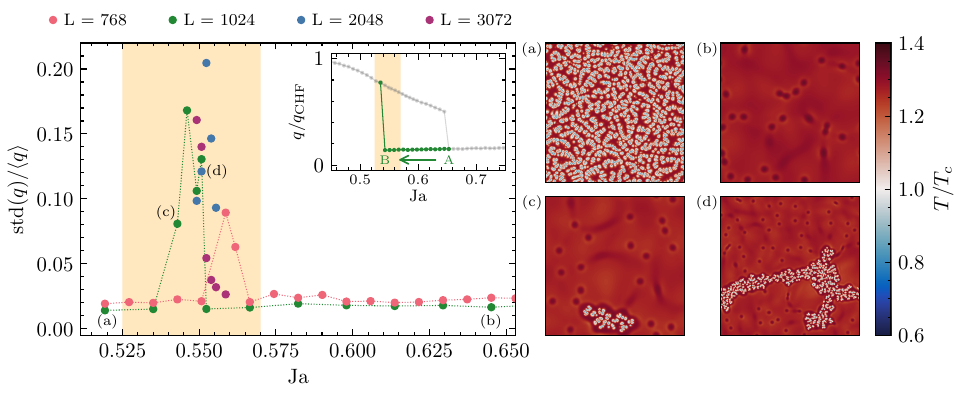}
    \caption{ Formation of metastable states during the rewetting process. The figure shows the standard deviation of 
              the heat flux normalized by its time averaged mean value for various system size. 
              All simulations are initialized from a configuration in the film boiling regime (shown in the inset).
              Representative snapshots of the temperature field, normalized by the critical temperature, illustrate: 
              (a) fully nucleate boiling, (b) fully film boiling, and two metastable coexistence states 
              corresponding to large values of $\textrm{std}(q)/\langle q \rangle$: (c) an isolated nucleate-boiling patch 
              embedded in a film-boiling background, and (d) a filamentary structure.
            }\label{fig:large_system}
\end{figure*}

In what follows, we will focus our analysis on the transition regime. 

In Fig.~\ref{fig:overview}(b), we provide results for simulations in a system with size $L = 512$, 
highlighting the presence of the two distinct branches of the hysteresis in the boiling curve. 
The red points correspond to simulations initiated from the transitional boiling regime, with the Jakob number gradually increased. 
Conversely, the green points represent simulations starting from the film boiling regime, with the Jakob number 
gradually decreased. The data clearly shows that the transition between the two regimes occurs at different Jakob numbers.

In order to better understand the transition between nucleate and film boiling, we examine the two branches separately, 
starting with the one obtained by increasing $\rm{Ja}$.
Figure~\ref{fig:small_system_analysis}(a) shows examples of the time evolution of 
the average heat flux in simulations where the system transitions from nucleate to film boiling.
We consider 80 independent realizations, each initialized from uncorrelated snapshots of a simulation at 
$\rm{Ja} = 0.527$, and restarted at $\rm{Ja} = 0.653$ (points A and B in the inset).
For each single realization, we measure the crossover time $t^{*}$, defined as the time required for the heat flux to drop down
to within $10\%$ of its mean value in the film boiling regime. The probability density function (PDF) of $t^{*}$ is
reported in panel~(b), along with a fit to an exponential distribution from the data (black line). Moreover, the inset
shows the empirical cumulative distribution function (CDF) plotted against the theoretical exponential distribution,
confirming an excellent agreement. 
This analysis underlines that the transition process is well described by a memoryless, Markovian mechanism, 
with a constant probability of the system switching from transitional to film boiling at a given instant, independent of the dynamic history of the state.
From a physical point of view, this is consistent with a picture in which the dynamics near the transition 
are dominated by stochastic interfacial events such as the nucleation, coalescence, and collapse of vapor structures,
which can be observed in animations of the numerical simulations, and the energy landscape can be envisioned 
as a shallow potential barrier separating both states, with fluctuations triggering spontaneous escapes. 
In our numerical simulations, however, we only observed one-way transitions: within the explored parameter
range, the system consistently evolved nucleate boiling toward film boiling, but we never observed
the reverse transition from film boiling back to nucleate boiling. 
This asymmetry suggests that, once established, the film boiling state is strongly self-sustaining under 
the present conditions, due to the stabilizing effect of the continuous vapor layer that 
suppresses the liquid-solid contact necessary for nucleation. This self-sustaining mechanism gives rise to the hysteretic behavior.

We now turn to the second branch of the hysteresis loop, obtained by starting in the film boiling regime 
and gradually decreasing $\rm{Ja}$ until the system returns to nucleate boiling.
In this case, the transition is characterized by the sporadic emergence of macroscopic structures exhibiting
\emph{spatial phase coexistence}. In Fig.~\ref{fig:large_system} we provide a few examples in the proximity of the transition.
In panel (a) we plot the standard deviation of the heat flux, normalized by its time average.
The pronounced peaks correspond to cases where different spatially separated regions of the domain simultaneously sustain nucleate and film boiling.

\begin{figure}[htb]
    \centering
    \begin{overpic}[width=\columnwidth]{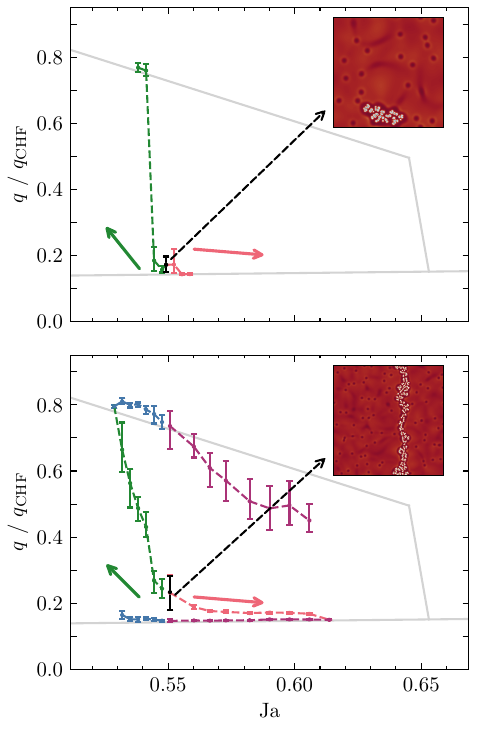}
     \put(-1,  98.0){\textbf{ (a) }}
     \put(-1,  49.0){\textbf{ (b) }}
    \end{overpic}        
    \caption{
    Impact of the phase coexistence morphology on the hysteresis dynamics. 
              (a) Average heat flux as a function of $\rm{Ja}$ starting from an isolated nucleate 
                  boiling patch within a film boiling background. 
              (b) Same, starting from a filamentary structure.
              In both panels red curves represent the case in which $\rm{Ja}$ is increased, and green the case in which
              is decreased. Gray lines show for reference the hysteresis curve obtained starting from configurations
              without phase coexistence shown in Fig.~\ref{fig:large_system}.
              In panel (b) we also show conditional averages of the heat flux, separately for the nucleate and film boiling 
              regions (blue and magenta lines). Dots show the heat-flux mean values, with error bars spanning the 
              10th-90th percentile range.
     }\label{fig:filaments}
\end{figure}

A more qualitative representation is given in the right panel of Fig.~\ref{fig:filaments} where we show snapshots 
of the temperature field (scaled by the critical temperature) near the bottom plate. 
For reference, the top-left and top-right images correspond to the endpoints of the hysteresis curves in the inset 
of the left panel, representing fully nucleate boiling and fully film boiling states, respectively. Light-colored regions
indicate colder zones where nucleation has occurred, reflecting the local heat consumption required to form vapor
bubbles. The bottom panels show two examples of spatial phase coexistence: on the bottom left (c), an isolated nucleate
boiling patch embedded in a film boiling background ($L=1024$); on the bottom right (d), a filamentary structure spanning
the domain ($L=2048$). We found these coexistence states to be long-lived metastable configurations, sustained for extended
periods rather than being mere transient fluctuations, see also End Matter. 

The formation of such metastable states is found to be inherently stochastic. Within a narrow region in the proximity of
the transition, the state of the system depends strongly on initial conditions: the system may evolve toward 
a metastable coexistence state, or directly into fully nucleate or stay in fully film boiling. 
The morphology and stability of coexisting regions are influenced by factors such as time scales,
the aspect ratio of the domain, the relative scale of bubbles to the system size, and boundary conditions.
A systematic study of these effects goes beyond the scope of this work, but we note that under the present parameters, 
the smallest system size at which phase coexistence is observed is $L = 768$.

We conclude by giving two examples of how these structures can significantly influence the hysteresis cycle.  
Figure~\ref{fig:filaments}~(a) shows that when starting from an isolated nucleate boiling patch within 
a film boiling background, transitions occur relatively quickly, as such patches can be easily nucleated or eliminated.  
In contrast, filamentary structures with a similar area proportion which span and partition the domain (panel~(b)) exhibit enhanced stability.  
These structures induce
large-scale circulation patterns reminiscent of Rayleigh-B\'{e}nard convection rolls \cite{Ahlers2009,lakkaraju2013heat}, leading to a feedback mechanism with the phase coexistence driving the large-scale circulation, 
and the circulation in turn stabilizing the coexistence state. 
Such configurations can be destroyed only by sufficiently strong perturbations or by moving far from the coexistence regime
(see End Matter, Fig.~\ref{fig:filament_dynamics_a}-\ref{fig:filament_dynamics_b} for an example).

{\it Conclusions.---}
In summary, based on extensive numerical simulations, we have provided clear evidence of hysteresis in the transition 
boiling regime, even under the idealized conditions of pool boiling on a flat surface kept at constant temperature. 
Our results demonstrate a clear asymmetry in the boiling transition depending on the direction of Jakob number variation. 
During heating, the transition from nucleate to film boiling is sharp and memoryless, consistent with a stochastic, 
Markovian mechanism dominated by interfacial events such as nucleation, coalescence, and collapse of vapor structures. 
In contrast, during cooling, the transition exhibits spatial phase coexistence, with macroscopic regions of nucleate 
and film boiling which may persist as long-lived metastable states, potentially delaying the rewetting process.

The stabilization of metastable states by large-scale convective flows represents a previously unrecognized mechanism in
boiling heat transfer. This finding suggests that flow patterns and thermal boundary conditions can be engineered to
either promote or suppress hysteresis, depending on the desired operational characteristics.

In future works we plan to move from the idealized conditions described in the present work and 
analyze the dependence of phase coexistence in the presence of  surface roughness, fluid properties, 
non-isothermal wall conditions and time-varying boundary conditions.

\begin{acknowledgments}
\medskip
This study was supported by the National Natural Science Foundation of China (Grant No. 52506104) and the HPC-Europe computational facilities and allocation on the MareNostrum 5 supercomputer (BSC, Spain) 
through grant EHPC-EXT-2023E02-035.
A.G. gratefully acknowledges the support of the U.S. Department of Energy through the LANL/LDRD Program under project
number 20240740PRD1 and the Center for Non-Linear Studies for this work.
\end{acknowledgments}

\bibliography{biblio}

\newpage
\onecolumngrid
\vspace{5mm}
\begin{center}
    \large \textbf{End Matter}
\end{center}

\twocolumngrid

\subsection*{Numerical method}
In this work,  the flow field is solved by a cascaded lattice Boltzmann method (CLBM) based on the D3Q19 lattice coupled with the pseudopotential model. In the CLBM, the collision is based on the central moments of the discrete distribution functions $f_i$, which are defined as~\cite{geier2006cascaded}:
\begin{equation}
    {\rm{ }}{{\tilde k}_{mnp}} = \left\langle {{f_i}\left| {{{({e_{ix}} - {u_x})}^m}{{({e_{iy}} - {u_y})}^n}{{({e_{iz}} - {u_z})}^p}} \right.} \right\rangle, 
\end{equation}
where ${\left|  \cdot  \right\rangle }$ denotes a nineteen-dimensional column vector and $ {{\bf{e}}_i} = [\left| {{e_{ix}}} \right\rangle ,\left| {{e_{iy}}} \right\rangle ,\left| {{e_{iz}}} \right\rangle ]  $ ($ i = 0,1,...,18 $) is the discrete velocity. A set of natural central moments is used, in the ascending order of ($ m+n+p $)~\cite{fei2018three}:
\begin{equation}
\begin{array}{l} 
\left| {{{\tilde T}_i}} \right\rangle  = [{{\tilde k}_{000}},{{\tilde k}_{100}},{{\tilde k}_{010}},{{\tilde k}_{001}},{{\tilde k}_{110}},{{\tilde k}_{101}},{{\tilde k}_{011}},{{\tilde k}_{200}},{{\tilde k}_{020}},\\{{\tilde k}_{002}},
{{\tilde k}_{120}},{{\tilde k}_{102}},{{\tilde k}_{210}},  
{{\tilde k}_{201}},{{\tilde k}_{012}},{{\tilde k}_{021}},{{\tilde k}_{220}},{{\tilde k}_{202}},{{\tilde k}_{022}}{]^ \top }. \\ 
\end{array}
\end{equation}
Before the collision, we first execute the transformation from the discrete velocity space to the central moments space by, ${\rm{ }}\left| {{{\tilde T}_i}} \right\rangle  = {\bf{N}}\left| {{T_i}} \right\rangle  = {\bf{NM}}\left| {{f_i}}\right\rangle$, where $ {\bf {M}} $ is a transformation matrix (from $\left| {{f_i}} \right\rangle$ to raw moments $\left| {{T_i}} \right\rangle$) and $ {\bf {N }}$ is a shift matrix (from raw moments to central moments). Then the central moments are relaxed to their equilibria independently, giving the post-collision states:
\begin{equation}
\left| {\tilde T_i^*} \right\rangle  = \left| {{{\tilde T}_i}} \right\rangle  - {\bf{S}}\left( {\left| {{{\tilde T}_i}} \right\rangle  - \left| {\tilde T_i^{eq}} \right\rangle } \right) + ({\bf{I}} - {\bf{S}}/2)\delta t\left| {{C_i}} \right\rangle, 
\end{equation}
where  $ \left| {{C_i}} \right\rangle  $ incorporates the effects of the force field into central moment space, and the relaxation matrix $ {\bf{S}} $ includes the relaxation rates for different central moments. The equalibrium central moments are defined as the continuous central moments of the Maxwell–Boltzmann distribution in the continuous velocity space, namely $\left| {\tilde T_i^{eq}} \right\rangle  = {[\rho ,0,0,0,0,0,0,\rho c_s^2,\rho c_s^2,\rho c_s^2,0,0,0,0,0,0,\rho c_s^4,\rho c_s^4,\rho c_s^4]^ \top}$, with the lattice sound speed $c_s=1/\sqrt{3}$.

The pseudopotential model is used to model the phase change dynamics, in which the interaction force among fluid particles is defined as~\cite{shan1993lattice,shan1994simulation},
\begin{equation}\label{e7-13}
{{\bf{F}}_{{\rm{int}}}} =  - G\psi ({\bf{x}})\sum\limits_i {w({{\left| {{{\bf{e}}_i}} \right|}^2})} \psi ({\bf{x}} + {{\bf{e}}_i}\delta t, t){{\bf{e}}_i},
\end{equation}
where $ G $ controls the strength of the interaction force and $ \psi $ is a density-based pseudopotential function. The weights are given as  $ w (0) = 1$, $ w (1) = 1/6$ and $ w (2) = 1/12$. To incorporate the realistic equation of state, the pseudopotential function can be chosen as~\cite{he2002thermodynamic,yuan2006equations}, $\psi  = \sqrt {2({p_{EOS}} - \rho c_s^2)/G{c^2}}$, where $ G=-1 $ is used to guarantee the term inside the square root is positive. In addition, due to the density contrast between the liquid and vapor phase, the system is affected by the buoyancy force, ${{\bf{F}}_b} =  - (\rho  - {\rho _{ave}})g{\bf{k}}$, where $\bf{k}$ is the vertical unit vector. As a result, the total force acting on the system is $ {\bf{F}} = {{\bf{F}}_{{\rm{int}}}} + {{\bf{F}}_b} $, whose effects are incorporated by the following forcing terms~\cite{linlin-pof-2020}:
\begin{equation}\label{eq9}
\begin{array}{l}
\left| {{C_i}} \right\rangle  = [0,{F_x},{F_y},{F_z},0,0,0,\eta ,\eta ,\eta ,{F_x}c_s^2,\\
{F_x}c_s^2,{F_y}c_s^2,{F_z}c_s^2,{F_y}c_s^2,{F_z}c_s^2,0,0,0{]^ \top },
\end{array}
\end{equation}
where $\eta  = (2\sigma {\left| {{{\bf{F}}_{{\rm{int}}}}} \right|^2})/[{\psi ^2}(s_e^{ - 1} - 0.5)\delta t]$ is a correction term to restore the thermodynamic inconsistency of the pseudopotential model, and $\sigma$ is a tunable parameter. After the collision step, the post-collision distributions are reconstructed by $\left| {f_i^*} \right\rangle  = {{\bf{M}}^{ - 1}}{{\bf{N}}^{ - 1}}\left| {\tilde T_i^*} \right\rangle $, followed by the free streaming in the discrete velocity space, ${f_i}({\bf{x}} + {{\bf{e}}_i}\delta t,t + \delta t) = f_i^*({\bf{x}},t)$. Then, the hydrodynamic variables are updated by:
\begin{equation}
\rho  = \sum\limits_i {{f_i}},~~~{\rm{   }}\rho {\bf{u}} = \sum\limits_i {{f_i}{{\bf{e}}_i}}  + \delta t{\bf{F}}/2.
\end{equation}
Using the Chapman-Enskog analysis, Eq. (\ref{eq:ns}) can be recovered in the low-Mach number limit, with viscosities related to the relaxation rates of second-order moments, $\nu  = (1/{s_2} - 0.5)c_s^2\delta t$ and  $ \xi  = 2/3(1/{s_{2b}} - 0.5)c_s^2\delta t $. The effective force field is given by:
\begin{equation}
    \bm{\Xi} = \bm{F} 
    - 2 G^2 c^4 \sigma \bm{\nabla} \cdot \left( |\bm{\nabla} \psi|^2 \mathbf{I}, \right),
\end{equation}
where the second term is induced by $\eta$ in Eq. (\ref{eq9}), which is used to adjust the mechanical stability solution of the system to match the thermodynamic theory~\cite{li2013lattice}. 

Following previous works~\cite{li2015lattice,linlin-pof-2020}, we set
$ \varpi=0.344 $, $a=2/49$, $b=2/21$ and $R=1$ in the equation of state, leading to the critical temperature and pressure as $T_c = 0.07292$ and  $p_c = 0.05957$, respectively. In the simulations, the saturation temperature is set as $ {T_s}= 0.86{T_c}  $, and $ \sigma=0.1  $ is used to achieve the thermodynamic consistent saturated densities ($ {\rho _l} = 6.4989 $ and $ {\rho _v} = 0.3797 $). The relaxation rates for the higher-order moments are set to be 1.2.

\begin{figure}[htb]
    \centering
    \begin{overpic}[width=0.85\columnwidth]{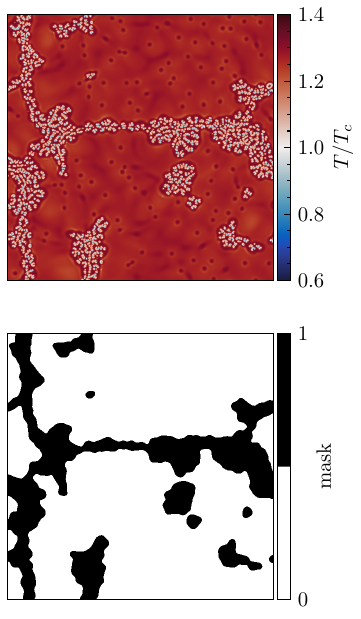}
     \put(-2, 100.0){\textbf{ (a) }}
     \put(-2,  49.0){\textbf{ (b) }}
    \end{overpic}
    \caption{ Example of the definition of a binary mask use to partition the system among regions in
              film/nucleate boiling state, used to compute conditional averages of the heat flux.
              Snapshot taken from a simulation with $L = 3072$, $\mathrm{Ja} = 0.55$.
            }\label{fig:masking}
\end{figure}

\subsection*{Calculation of heat-flux via conditional averages}

In the main text we have reported time averaged values of the heat flux as a function of the Jakob number.
We measure the heat flux at the bottom plate for a generic time step as
\begin{equation}\label{eq:hf}
  q(t) = - \frac{\lambda}{L^2} \int_0^L \int_0^L \left. \frac{\partial T}{\partial z} \right|_{z=0} \, dx \, dy ,
\end{equation}
where $\lambda$ is the thermal conductivity. 

In Fig.~\ref{fig:filaments}, where metastable states display phase coexistence, we have also reported averages 
of the heat-flux conditioning on whether specific areas are in the film or nucleate boiling state. 
To achieve this, we first analyze the spatial distribution of the temperature field near the bottom plate. 
Since bubble nucleation requires energy transfer for vapor formation, bubble-forming regions are systematically 
found at temperatures below the critical temperature $T_c$ (Fig.~\ref{fig:masking}a). 
We therefore construct an initial binary field by retaining only the subcritical temperatures. 
This binary field is convolved with a Gaussian kernel to smooth the data, bridging small gaps and connecting 
nearby subcritical regions while filtering out isolated patches arising from noise or transient fluctuations.
Finally, a binary mask is obtained by applying a threshold to the normalized smoothed field, where regions above
threshold are identified as coherent subcritical domains (Fig.~\ref{fig:masking}b). 
This mask enables the computation of spatially-conditioned ensemble averages.

\subsection*{Examples of time dynamics of filamentary structures}

\begin{figure*}[htb]
    \centering
    \includegraphics[width=0.99\linewidth]{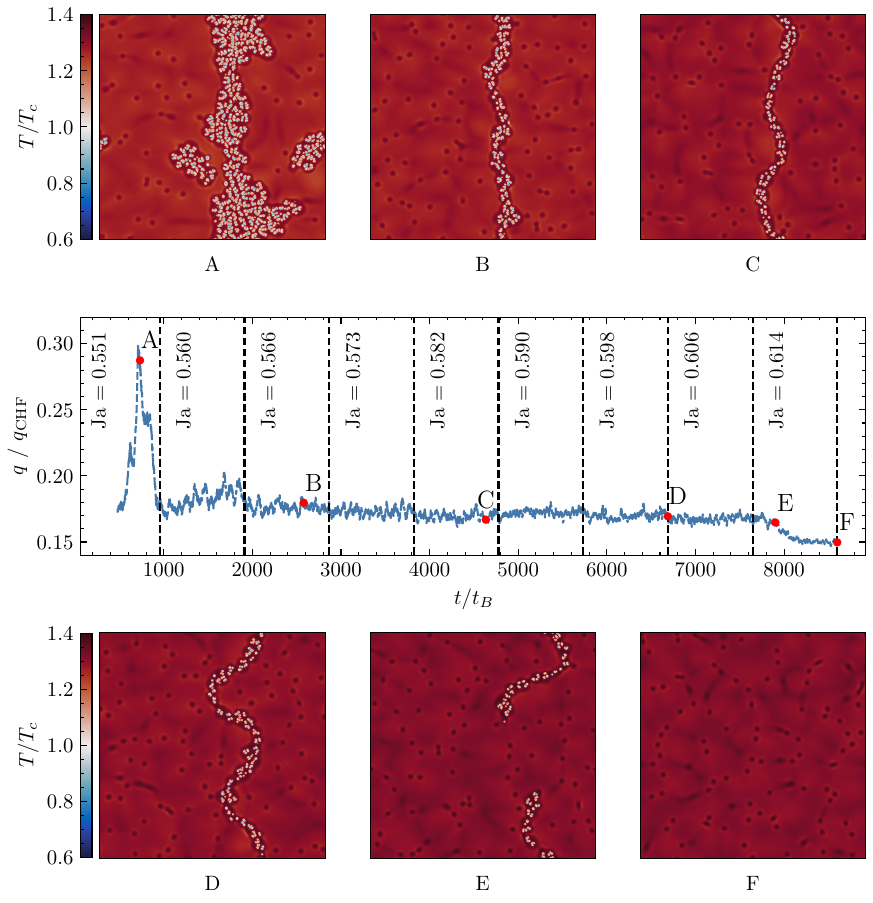}
    \caption{ Time evolution of the filamentary structures dynamically increasing the Jakob number. 
              The figure show the time evolution of filamentary structures highlighted in Fig.~\ref{fig:filaments}b. 
              The central panel displays the instantaneous heat flux as a function of time, 
              normalized by the characteristic buoyancy time $t_{B} = H/v_{B}$. 
              Dotted vertical lines indicate the times when the bottom-plate temperature is increased.
            }\label{fig:filament_dynamics_a}
\end{figure*}

\begin{figure*}[htb]
    \centering
    \includegraphics[width=0.99\linewidth]{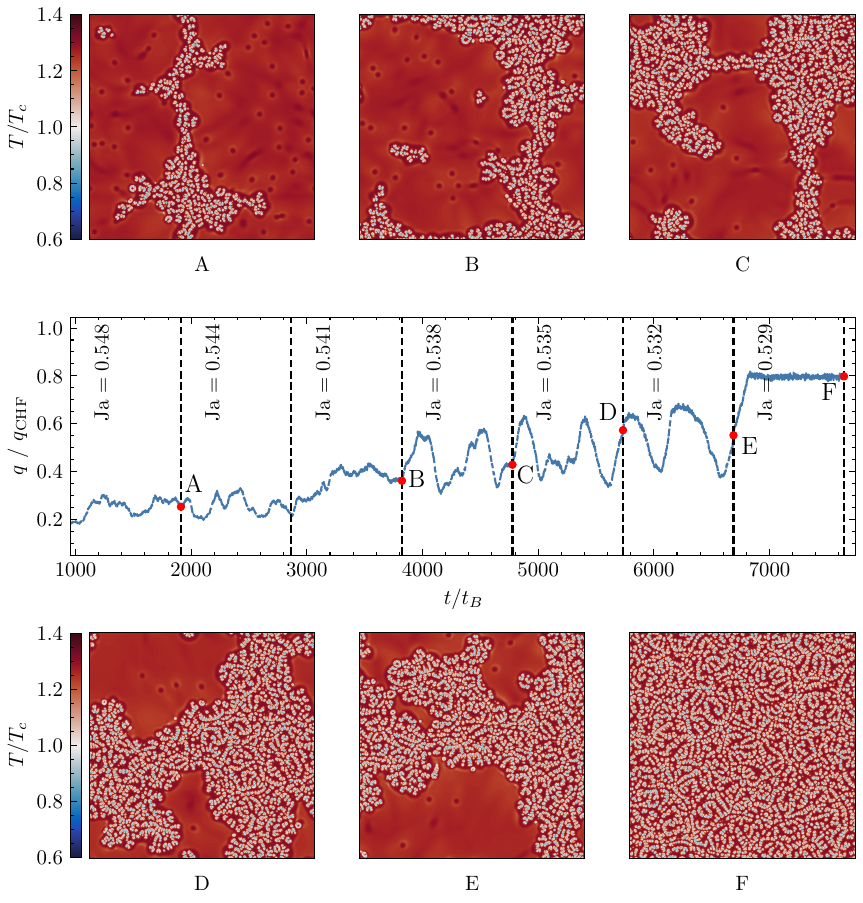}
    \caption{ Time evolution of the filamentary structures dynamically decreasing the Jakob number. 
              The figure show the time evolution of filamentary structures highlighted in Fig.~\ref{fig:filaments}b. 
              The central panel displays the instantaneous heat flux as a function of time, 
              normalized by the characteristic buoyancy time $t_{B} = H/v_{B}$. 
              Dotted vertical lines indicate the times when the bottom-plate temperature is decreased.
     }\label{fig:filament_dynamics_b}
\end{figure*}

We conclude by providing a more detailed insight in the dynamics experiences by the filamentary structures exposed in
Fig.~\ref{fig:filaments}b, showing the time evolution when increasing (Fig.~\ref{fig:filament_dynamics_a}) and
decreasing (Fig.~\ref{fig:filament_dynamics_b}) the Jakob number.
In both figures we show in the central panel the instantaneous value of the heat-flux (Eq.~\ref{eq:hf}) as a function
of time (in units of the characteristic buoyancy time $t_{B} = H/v_{B}$). Dotted vertical lines show the time instance
at which the temperature at the bottom plate is increased (or decreased). For each figure we also show 6 representative 
shapshots of the temperature field near the bottom plate, labeled in the heat flux plot with letters.
In Fig.~\ref{fig:filament_dynamics_a} (corresponding to the red curve in Fig.~\ref{fig:filaments}), we observe that
the phase coexistence (due to periodic boundary conditions) partitions the domain in two, leading to the creation of
large-scale circulation. When the temperature increase is sufficient to break the filamentary structure the system
transitions to purely film boiling state.

Fig.~\ref{fig:filament_dynamics_b}, corresponding instead to the green curve in Fig.~\ref{fig:filaments}, shows instead
the transition towards a purely nucleate boiling state obtained decreasing the temperature at the bottom plate.

\newpage

\end{document}